\documentclass[%
 reprint,
 amsmath,amssymb,
 aps,
 prl,
]{revtex4-1}
\newcommand{\avg}[1]{\langle #1 \rangle} 
\usepackage{physics}
\usepackage{amsmath}
\usepackage{mathrsfs}
\usepackage[font=small,labelfont=bf]{caption}
\usepackage{graphicx}
\usepackage{dcolumn}
\usepackage{bm}
\usepackage{bbm}
\usepackage{bigints}
\usepackage{graphicx}
\usepackage[percent]{overpic}
\usepackage{subfig}

\begin{document}


\title{Comparing experiments on quantum traversal time with the predictions of a space-time-symmetric formalism}

\author{Ricardo Ximenes}%
\author{Fernando Parisio}%
\author{Eduardo O. Dias}%
 \email{eduardodias@df.ufpe.br}
\affiliation{%
Departamento de Fisica, Universidade Federal de Pernambuco, Recife, Pernambuco 50670-901, Brazil
}%

\begin{abstract}
The question of how long a particle takes to pass through a
potential barrier is still a controversial topic in quantum mechanics.
Arguably, the main theoretical problem in obtaining estimates for
measurable times is the fact that previously defined time operators, that
remained within the borders of standard quantum mechanics,
present some kind of pathology. Recently, a time operator acting on an additional Hilbert space has been shown
to support both Hermiticity and canonical relation with an energy observable. The theory is built in a framework
which treats space and time as symmetrically as possible, in the nonrelativistic
regime. In this work, we use this formalism to derive a closed analytic expression for the
traversal time of a quantum particle impinging on a constant-potential barrier. We test our
theory in the specific experimental scenario of a realization by Rafagni et al [Appl. Phys. Lett. {\bf 58}, 774 (1991)].
The proposed approach displays a much better performance in comparison with the B\"uttiker-Landauer and the phase-time approximations.

\end{abstract}

\pacs{Valid PACS appear here}\maketitle


In quantum mechanics (QM) time is a label that we can
choose with arbitrary precision to analyze the state of a
system. Thus, QM does not directly take into account
the fact that we do not know in advance the exact instant at which the system
is measured. This is one of the main motivations for the several approaches to define
time operators, which would immediately bring this fundamental temporal ``fuzziness''
into the formalism. It is well known, however, that by trying to accommodate
time operators in the traditional framework of QM, both, conceptual and
mathematical problems arise, for instance, either the loss of Hermiticity or the non-canonical
commutation relation with an energy operator.

Recently, some of us proposed a space-time-symmetric extension of QM,
where the uncertainty about the moment of observation emerges
naturally ~\cite{DiasPar}.
The roles of time and space may be interchanged: depending on how the experimentalist
is measuring, the position of a particle may become a parameter,
whereas the measurement time is elevated to the status of an operator.
In these circumstances the time at which a measurement takes place is inherently
probabilistic and, importantly, the time operator so defined does not suffer from either lack Hermiticity or non-canonical
relation with energy.

Within the problems involving quantum time, those related to tunnelling are the most appealing, since they stand as one of the boldly nonclassical effects in QM.
Since MacColl's work back in 1932~\cite{mac} questions such as ``how long does
a particle take to traverse a spatial barrier?'' have been extensively studied, especially in the last few decades~\cite{But,Hauge1,Brouard,Hauge2,Landauer,Jonson,Olkho,Landmart}. These works were carried out from contrasting points of view, including quantum ``clocks'', Bohmian QM, and through path integrals.
The vast majority of the models try to describe time measurements by using a formalism inside the sphere of traditional QM. For instance, those models that try to predict traversal times do not present a satisfactory agreement with experimental data (see Ref.~\cite{ranfa}). Under these circumstances, in this work we deal with the traversal-time problem (which encompasses the tunnelling regime) by following the formalism proposed in Ref.~\cite{DiasPar}.  Then, we compare our results with experimental data from an optical realization~\cite{ranfa} that simulates the problems of traversal and tunnelling times, being formally equivalent to them.

Due to a close analogy between electromagnetic wave propagation and quantum particle dynamics, in the 1990's traversal times (also called delay-times) started to be investigated via optical experiments.
Here we compare the results predicted by the space-time-symmetric formalism proposed in \cite{DiasPar} with the experimental data of the delay-time measurements in narrowed waveguides~\cite{Mugran,ranfa}. We also compare our results with some of the most known analytical models for traversal time: the phase-time (PT) approach and the
B\"uttiker-Landauer (BL) model~\cite{But,Mugasala,Mugarush}. We show that our formalism presents a considerable improvement in describing the experimental data in comparison to the latter models.
%
%

We begin by presenting a brief summary of some results obtained in Ref.~{\cite{DiasPar}. A possible, though uncommon, interpretation of $|\langle x|\psi(t)\rangle|^2~dx=|\psi(x,t)|^2~dx$, within standard quantum mechanics, is that it represents the probability of observing a particle between $x$ and $x+dx$, \emph{given that} the time of the measurement is $t$. This interpretation emphasizes the conditional role of the parameter $t$ in the Schr\"odinger wave function. For this reason we set $\psi(x,t) \equiv \psi(x|t)$, hereafter.
Also, given the symmetry of the Bayes' rule, it is very natural to consider the ``mirror'' wave function $\phi(t|x)$, associated with the probability of measuring a particle in the time window
$[t,t+dt]$, {\it given that} the detection occurred at position $x$~\cite{DiasPar}. Explicitly, Bayes' rule relates these two functions as
\begin{eqnarray}
\label{P1}
P(x,t)dxdt=|\psi(x|t)|^2f(t)dxdt
=|\phi(t|x)|^2g(x)dxdt,\nonumber\\
\end{eqnarray}
where, $P(x,t)$ is the joint probability distribution, and $f(t)$ [$g(x)$] is the probability density of finding the particle at the instant $t$ [position $x$], regardless of the position $x$ [instant $t$] of the measurement. It is important to understand that the function $f(t)$ [$g(x)$] cannot be obtained solely through the Sch\"odinger state $|\psi(t)\rangle$. The distributions $f(t)$ and $P(x,t)$ correspond to new probability densities that are essential to describe data when position and time are measured simultaneously. For a more detailed discussion see~\cite{DiasPar}.
Note that in the second part of Eq.~(\ref{P1}) $x$ and $t$ play opposite roles in comparison with those played in Schr\"odinger QM. Here $t$ must be seen as the eigenvalue of a temporal Hermitian observable that will be introduced later, and $x$ is a continuous parameter that we can choose with arbitrary precision in order to evaluate the time dependence of $\phi(t|x)$.

Now let us look at the definition of the time operator elaborated in Ref.~\cite{DiasPar}. In standard QM, the state of a spinless particle in one dimension is given by $|\psi(t)\rangle \in \mathscr{H}_{X}$. The position of the particle is represented by the observable $\hat{X}$, so that $\hat{X}|x\rangle=x|x\rangle$, and the Sch\"odinger wave function is obtained by $\langle x|\psi(t)\rangle=\psi(x|t)$. In the extension proposed in Ref.\cite{DiasPar}, in order to turn $t$ into an operator, a new state $|\phi(x)\rangle$ and a time operator $\hat{T}$ are defined, related to a new Hilbert space $\mathscr{H}_{T}$. The eigenvalue of the operator $\hat{T}$ represents the time at which the system is observed, and by analogy, we should have $\hat{T}|t\rangle=t|t\rangle$. Therefore, the space-conditional wave function is given by $\langle t|\phi(x)\rangle=\phi(t|x)$. Moreover, in Ref.~\cite{DiasPar} a ``dynamic'' equation that governs the way $|\phi(x)\rangle$ evolves {\it in space} is proposed. In these new circumstances, temporal predictions should be made by states belonging to $\mathscr{H}_{T}$ instead of $\mathscr{H}_{X}$. Since in traversal time experiments the detector is fixed at a position after the barrier, it is natural to conclude that the wave function $\phi(t|x)$, describing spatially-conditioned temporal distributions, should predict the experimental outcomes.

The solution of the dynamic equation of $\phi(t|x)$ for the free particle case is obtained in Ref.~\cite{DiasPar}, and the corresponding probability density reads
\begin{multline}\label{rho}
\rho(t|x)=\frac{\hbar}{2\pi m} \abs{  \int_{0}^{\infty}\dd{k} \sqrt{k} C^{+}_{k} e^{ikx/\hbar-iE_{k}t/\hbar}}^2  \\ + \frac{\hbar}{2\pi m}\abs{  \int_{0}^{\infty}\dd{k} \sqrt{k} C^{-}_{k} e^{-ikx/\hbar-iE_{k}t/\hbar}}^2,
\end{multline}
where $C^{\pm}_k$ is the probability amplitude of the particle being measured with momentum $\pm  \hbar k$. As discussed above, Eq.~(\ref{rho}) is the probability amplitude of a temporal measurement at the instant $t$ \emph{given that} the particle has been detected at the position $x$. We will use Eq.~(\ref{rho}) to make temporal predictions for traversal-time experiments.
%
%
We proceed by defining expectation values of the time operator ${\hat T}$:
\begin{equation}\label{T1}
\avg{\hat T}(x)\equiv \frac{\mel{\phi(x)}{\hat{T}}{\phi(x)}}{\braket{\phi(x)}}=\frac{\bigintss_{-\infty}^{\infty} \dd{t}~t~\rho(t|x)}{\bigintss_{-\infty}^{\infty} \dd{t}\rho(t|x)},
\end{equation}
where we used the closure relation $\mathbbm{1}=\int dt|t\rangle\langle t|$.
Eq.~(\ref{T1}) defines the average of time measurements carried out at the position $x$.

We point out that the dynamic equation of the function $\phi(t|x)$, and its solution for the free particle case are deduced in Ref.~\cite{DiasPar} irrespective of the dynamics of the Sch\"odinger wave function $\psi(x|t)$. For this reason, it is important to note that, at a first glance, the coefficients $C_k^{\pm}$ of Eq.~(\ref{rho}) are unrelated to the coefficients $\tilde{C}_k$ of $\psi(x|t)$ for the free particle situation:
$\psi(x|t)=(2\pi \hbar)^{-1/2} \int_{-\infty}^{\infty}\dd{k} \tilde{C}_{k} e^{ikx-iE_k t/\hbar}$.
In addition, it should be noted that these coefficients are related to different experimental situations: the coefficients of $\psi(x|t)$ [$\phi(t|x)$] describe experimental results where time (position) is a conditional parameter.
However, by inspecting Bayes' rule, we verify that these two coefficients {\it must} be linked through relation~(\ref{P1}).
In addition, it is worth observing that $C_k^{\pm}$ and $\tilde{C}_k$ have similar interpretations which are related to the probability amplitude of the particle having momentum $\hbar k$. Thus, the simplest working hypothesis is $C_{k}^{\pm}=\Theta(\pm k) \tilde{C}_{k}$.
Note that Eq.~(\ref{rho}) can be written as
\begin{eqnarray}\label{d1}
\rho(t|x)=\frac{\hbar}{2\pi m}\sum_{r=\pm}  \int_{0}^{\infty}\dd{k} \int_{0}^{\infty}\dd{k'} \sqrt{kk'} C^{r}_{k}  {C}^{r^{*}}_{k'}
 \nonumber\\ \times e^{ir(k-k')x} e^{-i\hbar(k^2-k'^2) t/(2m)},
\end{eqnarray}
where $\ast$ denotes complex conjugation. To evaluate Eq.~(\ref{T1}), let us begin with the integral
\begin{eqnarray}\label{d2}
&&\int_{-\infty}^{\infty}\dd{t}~t~ \rho(t|x)=\frac{\hbar}{2\pi m}\sum_{r=\pm}  \int_{0}^{\infty}\dd{k} \int_{0}^{\infty}\dd{k'} \sqrt{kk'} C^{r}_{k}  {C}^{r^{*}}_{k'}  \nonumber\\
&&\times e^{ir(k-k')x}\int_{-\infty}^{\infty}\dd{t}~t~ e^{-i\hbar(k^2-k'^2) t/(2m)}\equiv I.
\end{eqnarray}
The integral in the variable $t$ in Eq.~(\ref{d2}) can be written as
$2\pi m/(i\hbar k') \partial_{k'}\delta\left[\hbar(k^2-k'^2)/(2m)\right]$, where $\partial_{k'}=\partial/\partial {k'}$ and $\delta(x)$ is the Dirac delta function. By using the relation $\delta\left[(\hbar(k^2-k'^2)/2m)\right]=(m/\hbar k)\left[\delta(k+k')+\delta(k-k')\right],$ and by noting that in Eq.~(\ref{d2}) $k,k'\geq 0$, the only contribution comes from $\delta(k-k')$.
The substitution of the resulting expression into Eq.~(\ref{d2}), the calculation of the integral in the variable $k'$, and the definition $\Gamma^{\pm}_{k}(x)\equiv C_{k}^{\pm}e^{\pm ikx}/\sqrt{k}$, lead to
\begin{equation}\label{int}
I=\frac{m}{i\hbar} \int_{0}^{\infty}\dd{k} ~\left[ {\Gamma}^{+^{*}}_{k}(x)~\dfrac{\partial \Gamma_k^+(x)}{\partial k}~+~{\Gamma}^{-^{*}}_{k}(x)~\dfrac{\partial \Gamma_k^-(x)}{\partial k}\right].
\end{equation}
For the normalization constant, similar calculations give
$\int_{-\infty}^{\infty}\dd{t} \rho(t|x)= \int_{0}^{\infty}\dd{k}\Big[|C^{+}_{k}|^2+|C^{-}_{k}|^2\Big]$.
Finally, we get
\begin{equation}\label{T3}
\avg{\hat T}(x)=\dfrac{\dfrac{m}{i\hbar} \bigintss_{0}^{\infty}  \dd{k} \left[\Gamma_k^{+^{*}}(x)\dfrac{\partial \Gamma_k^+(x)}{\partial k}+ \Gamma_{k}^{-^{*}}(x)\dfrac{\partial \Gamma_k^-(x)}{\partial k} \right]}{\bigintss_{0}^{\infty}\dd{k}\left[\abs{C_{k}^{+}}^2+\abs{C_{k}^{-}}^2  \right]}.
\end{equation}
Eq.~(\ref{T3}) is the expression for the expectation value of the time operator ${\hat T}$ evaluated at the position $x$ for the free particle. However, if one assumes $k=\sqrt{2m(E_k-V_0)}$ this expression still holds for a particle under the influence of a constant potential $V_0$.
\begin{figure*}[!ht]
    \subfloat{%
      \begin{overpic}[width=0.42\textwidth]{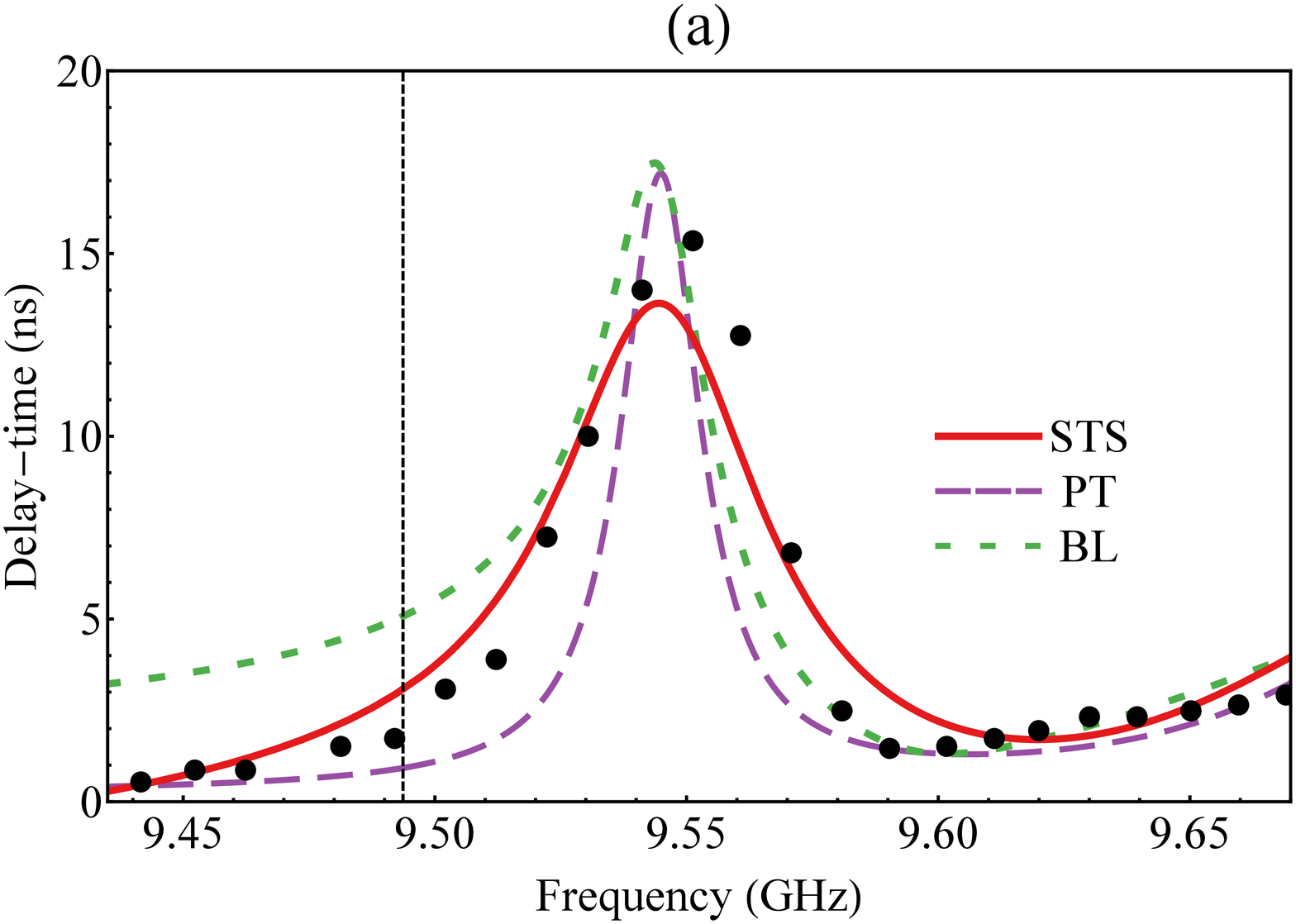}
        \put(65.8,42.7){\includegraphics[width=27\unitlength]{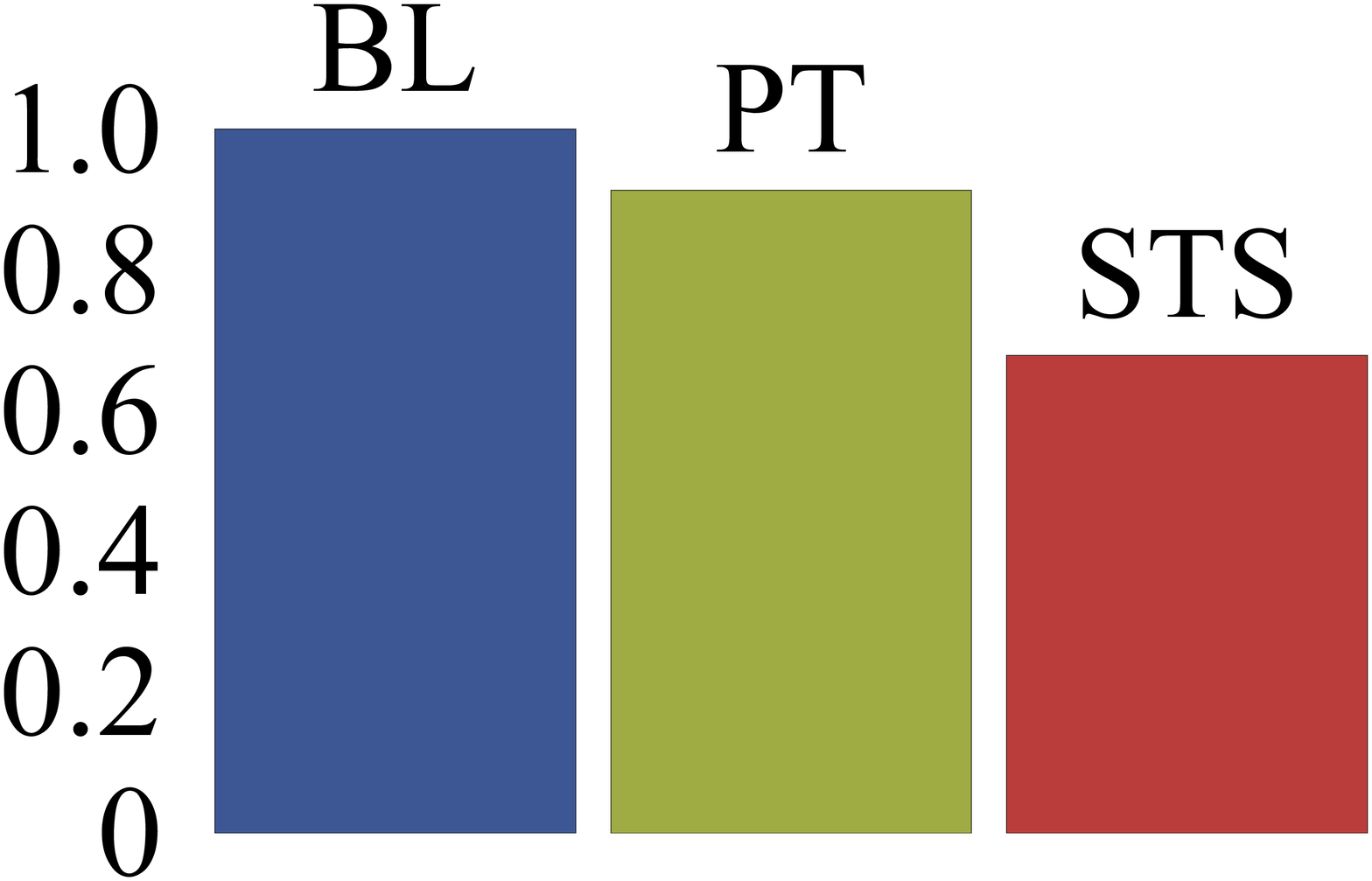}}
      \end{overpic}
    }
    \hspace{1cm}
      \subfloat{%
      \begin{overpic}[width=0.432\textwidth]{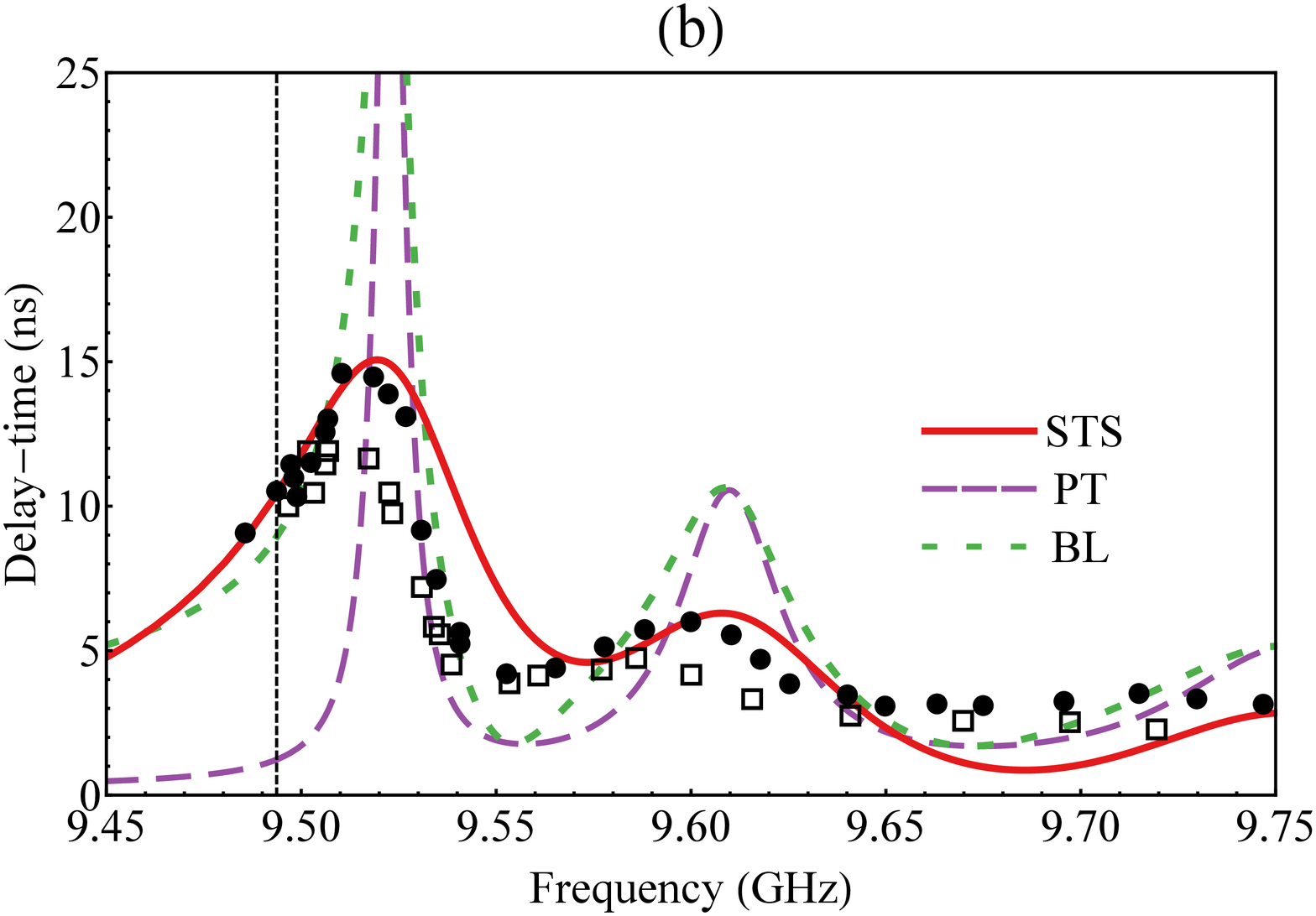}
        \put(63,41){\includegraphics[width=27\unitlength]{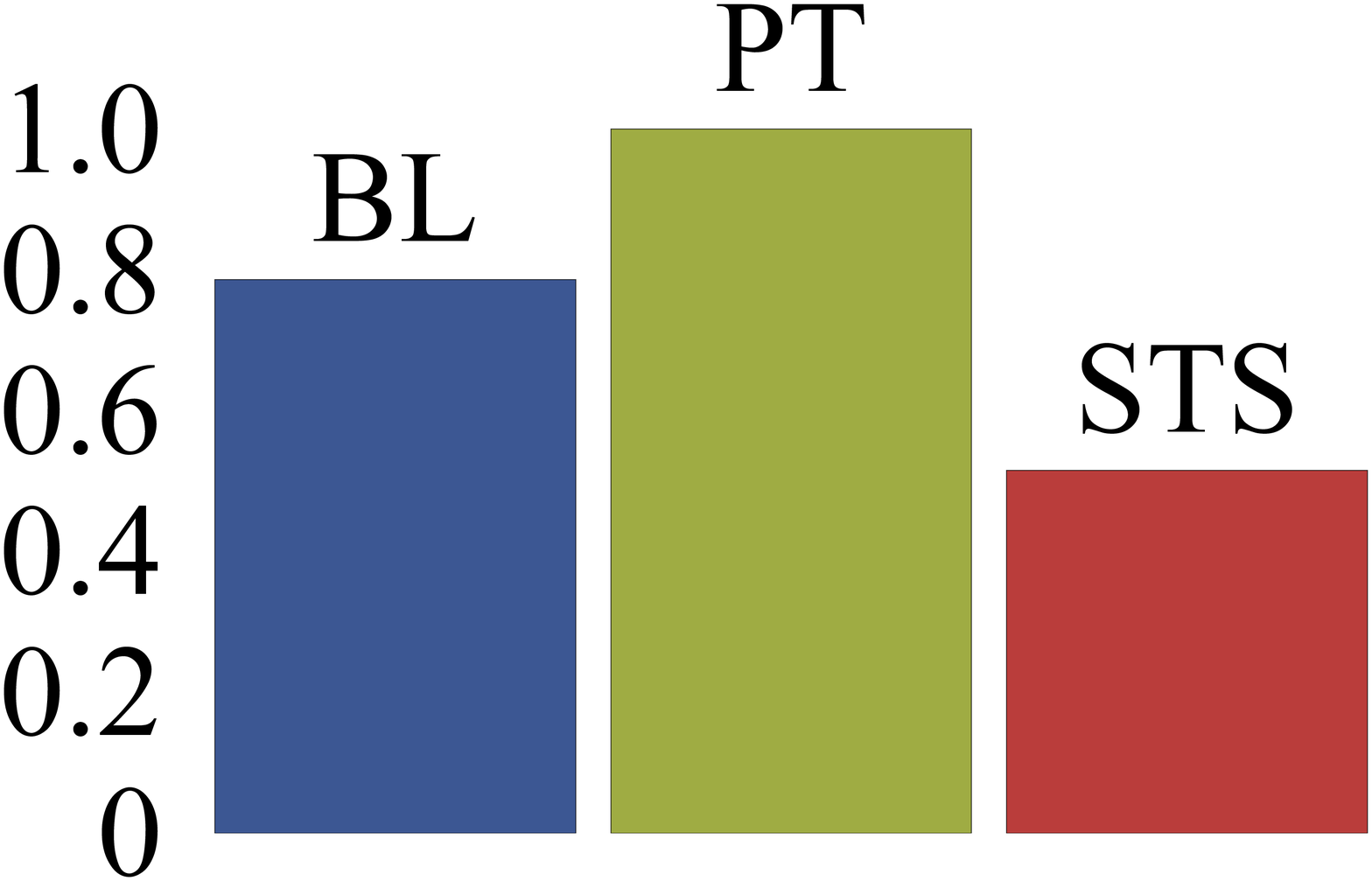}}
      \end{overpic}
    }
    \caption{(Color Online) Delay-time measurements for a ``potential barrier'' of length $L$ vs frequency. The solid curve represents Eq. (\ref{Ttrav2}), the long dashed curve represents the PT model, and the short dashed curve illustrates the BL model. (a) Potential barrier of length $L=15$ cm. A bandwidth $\Lambda= 30$ MHz is used for our model. The dots represent the experimental data obtained by Ref.~\cite{ranfa}. (b) Potential barrier of lenght $L=20$ cm. In this case,  the bandwidth is $\Lambda= 50$ MHz. The bullets and open squares represent two different runs of measurements ~\cite{Mugran,ranfa}. The vertical line represents the cut-off frequency.} \label{graf}
\end{figure*}
Let us use Eq.~(\ref{T3}) to calculate the mean value of the time for a particle crossing a potential barrier $V_0$  ($0<x<L$). Due to the space-conditional character of $\avg{\hat T}(x)$, and because the detectors in the optical experiment are fixed just before and after the barrier, we define the delay-time as
\begin{equation}\label{Ttrav}
T_{delay}(L)\equiv \avg{\hat{T}}(L)-\avg{\hat{T}}(0).
\end{equation}
To obtain $\avg{\hat{T}}(L)$, first we have to calculate $C_{k}^{\pm}$ for time measurements immediately after the barrier. Therefore, we need the stationary Schr\"odinger wave function in the region $x>L$, $\psi_{k}(x)\sim{\cal T}(k)e^{ikx}$, where ${\cal T}(k)$ is the transmission coefficient that is given by
\begin{equation}\label{8}
{\cal T}(k)=\frac{4k k_{1}e^{- i L (k-k_{1})}}{(k+k_{1})^2-e^{2iLk_{1}}(k-k_{1})^2}.
\end{equation}
Here, $k=\sqrt{2mE}$ and $k_{1}=\sqrt{2m(E-V_{0})}$ are the wave numbers outside and inside the barrier, respectively. So, for $x>L$
$C^{+}_k=A_{k}~{\cal T}(k)$ and $C_{k}^-=0$, where $A_k$ corresponds to the amplitude of the Schr\"odinger wave function of the incident wave packet in the barrier. Therefore, $\avg{\hat{T}}(L)$ can be written as
\begin{widetext}
\begin{equation}\label{Ttrav2}
\avg{\hat{T}}(L)= \dfrac{\dfrac{m}{i\hbar} \bigintss_{0}^{\infty}  \dd{k}~ \left (\dfrac{A_{k}}{\sqrt{k}} ~{\cal T}(k) e^{ikL}\right)^{*}~\dfrac{\partial}{\partial k}\left( \dfrac{A_{k}}{\sqrt{k}}~{\cal T}(k)e^{ikL} \right)}{\bigintss_{0}^{\infty}\dd{k}~\abs{A_{k}~{\cal T}(k)}^2}.
\end{equation}
\end{widetext}
Equation~(\ref{Ttrav2}) corresponds to our main analytical result. From now on, we will refer to Eq.~(\ref{Ttrav}) as the space-time-symmetric (STS) model.
%

The experiment \cite{Mugran,ranfa}, which we will address to test our predictions, consists of an $X$-band microwave circuit with a step narrowing waveguide in $P$ band whose length is given by $L$. The sections are $a'\times b'=7.9 \times 15.8 ~{\rm mm}^2$ in $P$ band, and $a\times b=10.16 \times 22.86~ {\rm mm}^2$ in $X$ band. The electromagnetic signals are sent to an oscilloscope with high resolution, and are recorded before and after the narrowing, and then the delay-time is measured. For more details about the setup, see Refs.~\cite{Mugran,ranfa}.

For the mode $TE_{0,1}$, the refractive index in the waveguide is $n=\{1-[\lambda/(2b)]^2\}^{1/2}$, with $\lambda$ representing the wavelength in free space. With this in mind, the rectangular barrier problem is equivalent to the optical circuit when the following replacements are made:
\begin{eqnarray}\label{kk1}
k&=&\frac{1}{\hbar}\sqrt{2mE}~\rightarrow~\frac{2\pi}{\lambda}n=\frac{2\pi}{c}(\nu^2-\nu_{out}^2)^{1/2}~,~~{\rm and} \nonumber\\ k_{1}&=&\frac{1}{\hbar}\sqrt{2m(E-V_0)}~\rightarrow~\frac{2\pi}{\lambda}n'=\frac{2\pi}{c}(\nu^2-\nu_{in}^2)^{1/2},\nonumber\\
\end{eqnarray}
where $k$ and $k_1$ are the wave numbers outside and inside the potential barrier respectively, and $\nu$ is the microwave source frequency. The constants $\nu_{in}$ and $\nu_{out}$ in the previous expression depend on the experimental parameters $b$ and $b'$ as follows,
$\nu_{in}=c/2b'$, $ \nu_{out}=c/2b$. Note that by substituting Eq.~(\ref{kk1}) into the dispersion relation for quantum waves $\hbar^2 k_1^2/(2m)=\hbar^2 k^2/(2m)+V_0$, we obtain the equivalent potential $V_0$ for the optical system $k_0\equiv \sqrt{2mV_0}/{\hbar}=2\pi/c(\nu_{in}^2-\nu_{out}^2)^{1/2}$. Finally, notice that the group velocity for quantum waves is $d\omega /dk=\hbar k/m$, whereas for electromagnetic waves in the waveguide we have $d\omega /dk=c^2k/\omega$. Therefore, in order to apply the equations of quantum mechanics to the waveguide problem, we have to make the substitution
$\hbar /m \rightarrow c^2/(2\pi \nu)$.
By substituting  \eqref{kk1} and the previous relations into \eqref{Ttrav2}, and defining $A_{\nu}\equiv\sqrt{\partial k/{\partial \nu}}A_{k}$, we have
\begin{eqnarray}\label{Ttrav3}
&&\avg{\hat{T}}(L)=\nonumber\\
&&\dfrac{\bigintss_{\nu_{out}}^{\infty}  \dd{\nu}~ \bigg({\cal T}(\nu)A_{\nu}e^{ik(\nu)L}\bigg)^*~\dfrac{\partial}{\partial \nu}\bigg({\cal T}(\nu)A_{\nu}e^{ik(\nu)L}\bigg)}{2\pi i \bigintss_{\nu_{out}}^{\infty}\dd{\nu}\abs{\cal T(\nu) A_{\nu}}^2}.\nonumber\\
\end{eqnarray}
We complete the analogy between the wave function of QM and the electromagnetic wave in the traversal time problem by associating the modulus square of the coefficients $A_{\nu}$ with the spectral line intensity of the source. The microwave signal is supplied by a klystron (Varian X-13) whose typical bandwidth is of the order of $30~{\rm MHz}$~\cite{klys}, with a central frequency of the order of 10 GHz. Therefore, we have an approximately monochromatic source. The typical line-width profile is Lorentzian, so we have
\begin{equation}\label{dist}
A_{\nu}=\frac{\sqrt{\Lambda/(2\pi)}e^{-i2\pi\nu t_{\mu}}}
{i(\nu+\nu_{\mu})+(\Lambda/2)}-\frac{\sqrt{\Lambda/(2\pi)}e^{-i2\pi\nu t_{\mu}}}{i(\nu-\nu_{\mu})-(\Lambda/2)},
\end{equation}
where $\Lambda$ is the scale parameter which specifies the half-width at half-maximum so that we consider $O(\Lambda)=30~{\rm MHz}$, and $\nu_{\mu}$ is the central frequency of the Lorentzian distribution. In Eq.~(\ref{dist}), $t_{\mu}=\ell/v_{\mu}$, with $\ell$ being the distance traveled by the electromagnetic pulse to arrive at the beginning of the barrier ($x=0$), and $v_{\mu}$ is the phase velocity of the mean frequency $\nu_{\mu}$. With this, note that by considering the phase factor $\exp{-i2\pi\nu t_{\mu}}$ in Eq.~(\ref{dist}) we already take into account the time $\avg{\hat{T}}(0)$ it takes for the pulse to reach the barrier. In these circumstances, we can rewrite the traversal time as $T_{delay}(L,\nu_{\mu})=\avg{T}(L,\nu_{\mu})$, with $A_{\nu}$ given by (\ref{dist}).

Now we turn our attention to the comparison of the predictions of STS model given by Eq.~(\ref{Ttrav3}) with the experimental results of Refs.~\cite{ranfa,Mugran}, as well as with the other theoretical models described in the supplemental material. To make the comparison with previous results easier, Fig.~\ref{graf} follows the guidelines used to represent similar plots in Refs.~\cite{ranfa,Mugran}: they illustrate how the traversal time (delay-time) varies as a function of the mean frequency $\nu_{\mu}$.

We consider typical values of the klystron bandwidth~\cite{klys}: in Fig.~\ref{graf}(a), we directly set $\Lambda=30~{\rm MHz}$. In this realization  $L=15~{\rm cm}$. We plot three theoretical curves: (i) STS [solid curve], BL [long-dashed line], and PT [short-dashed line]. The vertical dashed line represents the cut-off frequency, so that frequencies below this value refer to tunnelling times. The inset in this figure represents the residues, $\delta_k=N^{-1}\sqrt{\sum_i[y_i-f_k(x_i)]^2}\equiv N^{-1} \Delta_k$, where $(x_i,y_i)$ represent the experimental points and $f_k$ denotes the theoretical expression (BL, PT, STS), $k=1,2,3$, respectively. The largest residue is normalized to unity, $N=\max\{\Delta_1,\Delta_2,\Delta_3\}$.  The STS prediction presents the best performance while the BL approximation fails in describing the results in the left (tunnelling) part of Fig.~\ref{graf}(a). In this case the PT prediction is slightly better than BL.

In Fig.~\ref{graf}(b) we have $L=20~{\rm cm}$, while all other parameters are the same as in Fig.~\ref{graf}(a). In this case the analysis is more subtle because, according to the authors of ~\cite{ranfa,Mugran}, the difference between the amplitude of the previous theoretical models and the measured values is due to the neglecting of losses. In this case there were two runs (filed bullets and empty squares), which illustrates the variability of the results under the same experimental conditions. The inset depicts the residues referring to the bullets, which seems to present less losses. Since the description of the original experiment is vague in what concerns these attenuations, we follow the simplest possible path to take this effect into account. It is known that attenuation leads to a increase of the bandwidth~\cite{attenuation}, so we searched larger values of $\Lambda$, keeping it within tens of MHz, which best fits the experimental points. The obtained effective value is $\Lambda=50$ MHz. Again the STS predictions are much better than those by the BL and PT models. In this case the PT model presents the poorest performance.

%
%
The fact that a microscopic particle may traverse a potential barrier that would be classically insurmountable is one of the most emblematic phenomenon in quantum mechanics. In this work we applied a recently developed supra-quantum formalism to such a tunnelling problem. The referred theoretical framework relies on simple statistical reasonings and symmetry assumptions, and intends to equip quantum mechanics to deal with statistical-inference situations not embraced by the standard theory. In this particular case, this amounts to predict temporal probability distributions given that the position is fixed.

Within the theoretical framework proposed in \cite{DiasPar}, whenever position is fixed, it can be seen as a parameter. In addition, one can define an Hermitian time operator $\hat{T}$ satisfying canonical commutation relations with an energy observable. It then becomes natural to define the delay caused by the presence of a potential barrier, spatially localized
between $x=0$ and $x=L$, as $\langle\hat{T}\rangle(L)-\langle\hat{T}\rangle(0)$. Using this definition and standard information on a particular, well-known analog experiment \cite{Mugran,ranfa}, we were able to demonstrate the effectiveness of our approach. The experiments consisted in the study of the propagation of microwave electromagnetic radiation through a wave guide with a bottleneck, which is formally equivalent to a typical quantum tunnelling problem.  We compared our predictions with two well-stablished approximations, namely, Buttiker-Landauer and phase-time and our results show much closer agreement with the experimental points in two distinct experimental situations.
\acknowledgements
The authors thank Ant\^onio Azevedo and Fl\'avio Aguiar for helpful discussions on the klystron microwave source. CNPq and FACEPE (Brazilian Research Councils) are acknowledged for financial support. E. O. D. acknowledges financial support from FACEPE through PPP Project No. APQ-0800-1.05/14.

\end{document}